\documentclass[aps,english,prb,twocolumn,superscriptaddress]{revtex4}
\usepackage{graphicx}
\usepackage[left]{lineno}
\usepackage{gensymb}
\usepackage{amsmath}

\makeatletter
\def\p@figure{Fig.~}
\makeatother

\begin{document}

\title{Interfacial Charge Transfer and Electronic Structure Modulation in Ultrathin Graphene P3HT Hybrid Heterostructures}

\author{Yosra Mater}
\affiliation{Department of Physics, K\i r\i kkale University, K\i r\i kkale 71450, Turkey.}

\author{Salih Demirci}
\affiliation{Department of Physics, K\i r\i kkale University, K\i r\i kkale 71450, Turkey.}

\author{V. Ongun {\"O}z{\c{c}}elik} \email{ongun.ozcelik@sabanciuniv.edu}
\affiliation{Faculty of Engineering and Natural Sciences, Sabanci University, Istanbul, 34956 Turkey.}
\affiliation{Materials Science and Nano Engineering Program, Sabanci University, Istanbul, 34956 Turkey.}

\begin{abstract}
Ultrathin polymer/graphene heterostructures are promising materials for next-generation optoelectronic and photovoltaic technologies, while the influence of the polymer’s structural variation on interfacial charge transfer remains unclear. Here, using ab initio quantum mechanical calculations we show how different forms of Poly(3-hexylthiophene) (P3HT), a widely used organic semiconductor, interact  with graphene. We analyze the effects of molecular chain length, end-group termination, periodicity, and the distinction between ordered and random P3HT arrangements. For isolated P3HT, the band gap decreases with increasing chain length and layer thickness, while structural disorder leads to slightly larger gaps due to reduced electronic coupling. When P3HT is deposited on graphene, all configurations exhibit spontaneous charge transfer, with electrons accumulating on graphene and holes remaining in the polymer. This effect is significantly enhanced in ordered and fully periodic structures and is noticeably weaker in disordered ones. Charge density analyses further show that thicker and more ordered P3HT layers improve electron hole separation across the interface. Our results reveal how molecular structure governs charge transfer in P3HT/graphene heterojunctions and provide practical guidelines for designing high-efficiency polymer/graphene  photovoltaic devices.
\end{abstract}

\maketitle

\section{Introduction}

Ultrathin metal–organic hybrid heterostructures, combining an organic semiconducting polymer with a metallic conductive substrate, mark a recent advancement in flexible electronics, energy storage, and photovoltaic material design\cite{bae2019integration, huynh2002hybrid, chang2010high, ozcelik2020hybrid}. When a semiconducting polymer contacts a metallic substrate, electron-hole separation occurs within the heterostructure such that holes become localized in the organic polymer while electrons are directed to the metallic substrate, a process previously referred as spontaneous charge transfer \cite{xiang2017ultrafast}. This phenomenon critically impacts the heterostructure's optical and electronic behavior by perturbing the interfacial band structure and strengthening interlayer coupling through hybridization induced charge accumulation localized at the contact region \cite{braun2009energy}. The efficiency of interfacial charge transfer is strongly dependent on the atomic-scale registry at the junction, the electronic and chemical character of the interface, the polymer's adsorption geometry, and the degree of hybridization, all of which collectively determine the local band alignment, carrier injection barriers, and the resulting contact resistance in the device \cite{dag2008modeling,dag2010packing}. Therefore, a systematical understanding of these parameters and their impact on spontaneous charge transfer is essential for each interface type.

\begin{figure*}
	\includegraphics[width=15cm]{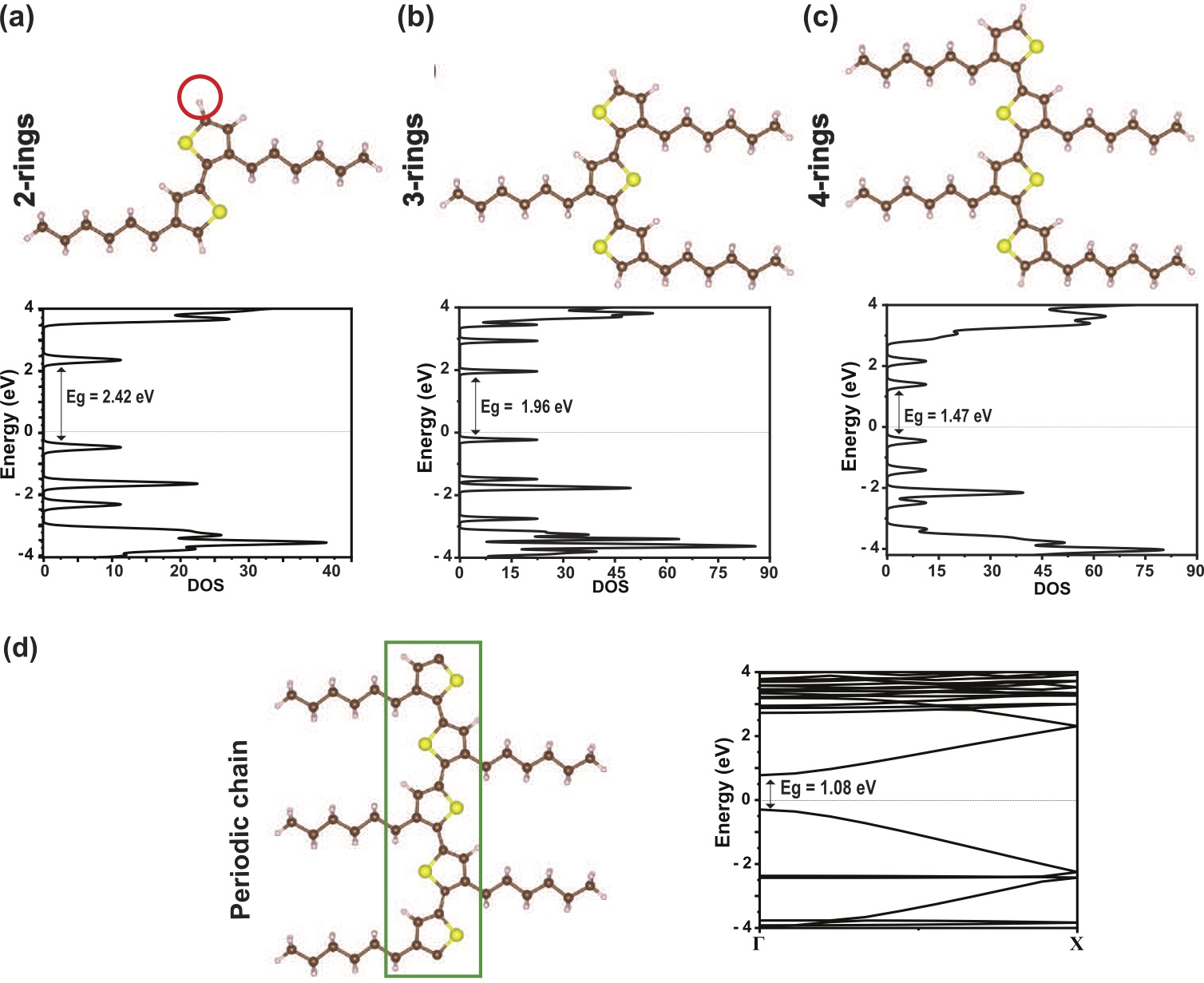}
	\caption{(a-c) Molecular structures and density of states (DOS) of finite P3HT molecules with  2, 3, and 4 thiophene rings. The H-termination of the thiophene ring is indicated with the circle in (a). (d) Regularly aligned periodic P3HT chain and its band structure . The Fermi level is aligned to zero.}
	\label{fig_molecule}
\end{figure*}

Poly(3-hexylthiophene) (P3HT) is a widely used conjugated polymer in organic solar cells and photovoltaic devices, valued for its high charge mobility, solution processability, and tunable optoelectronic properties \cite{liu2009photovoltaic, fratini2020charge, lucke2015p3htdefects, liu2011nanofibril, rudyak2020stacking, mardi2020thermoelectric}. On the other hand, graphene, known for its high charge mobility, mechanical flexibility, and large surface area, can function as an efficient two-dimensional(2D) substrate that enables rapid charge transport and robust interfacial coupling. These are two key features required for efficient photovoltaic systems\cite{huang2012graphene, kumar2021carbon, novoselov2012roadmap, yang2013graphene, weiss2012graphene, sun2011graphene}. Therefore, the integration of P3HT with graphene synergistically combines the complementary properties of both materials, leading to improved photovoltaic performance. Previous studies have shown that graphene can influence the crystallization of P3HT, resulting in different crystallite orientations compared to growth on conventional substrates like silicon. For instance, P3HT molecules deposited on graphene can exhibit a higher degree of $\pi$ - $\pi$ stacking perpendicular to the film plane, potentially enhancing vertical charge transport.\cite{che2018phototransistor, yu2010p3htgraft} Furthermore, strong photoluminescence quenching has been observed at the graphene - P3HT interface, which supports the presence of interfacial electron transfer \cite{skrypnychuk2015vertical}. Other works explored the use of solution-processed functionalized graphene in organic bulk heterojunction photovoltaic devices with P3HT, showing the dependence of photovoltaic performance on graphene content and morphology \cite{liu2009photovoltaic,liu2008grapheneacceptor,amalia2024first}.

Despite significant progress in understanding molecular P3HT on graphene, the impact of structural variations in periodically repeating P3HT molecules remains poorly understood. Such variations, which give rise to mono- and multilayer films with distinct regular and random P3HT order (ReP3HT and RaP3HT) can critically influence the interfacial band  level alignment, exciton dissociation efficiency, and charge transport in organic photo voltaic (OPV) devices. Understanding the electronic structures of these interfaces, along with the charge transfer mechanisms and degree of hybridization, is essential for the development of next-generation graphene-based OPV devices.

In this paper, we address this gap through systematic density functional theory (DFT) analyses, providing a comprehensive investigation of the electronic and structural properties, as well as the interfacial charge-transfer dynamics, in vertical 2D graphene/P3HT heterostructures. We focus on variations in P3HT configuration, starting from molecular P3HT and extending into the formation of monolayer to multilayers of P3HT, comparing regular and randomized polymer chain structures, and modeling systems from periodic chains to finite molecules with different end terminations. We aim to (i) characterize the electronic properties of free-standing P3HT in different structural forms, (ii) assess how these properties are altered upon integration with a graphene substrate, and (iii) investigate the interfacial charge-transfer mechanisms that govern the performance of P3HT/graphene based optoelectronic and photovoltaic materials.

\section{Methods}

The first-principles DFT calculations were performed using the Vienna \textit{ab initio} simulation package (VASP)~\cite{kresse1995abinitio,kresse1996efficient}. To describe the ion-electron interactions, projector augmented-wave (PAW) potentials were used~\cite{blochl2002paw}. The exchange-correlation energy was described using the generalized gradient approximation (GGA) within the Perdew-Burke-Ernzerhof (PBE) scheme~\cite{perdew1996gga}. Van der Waals interactions were included through the DFT-D3 method with Becke-Johnson damping~\cite{grimme2011d3}. The equilibrium atomic configurations were obtained by minimizing the total energy using the conjugate gradient method~\cite{Shewchuk1994AnIT}. The convergence criterion for the electronic self-consistency loop was set to $10^{-5}$~eV between two successive iterations. A plane-wave basis set with an energy cutoff of 520~eV was employed. For Brillouin-zone integration, a $\Gamma$-centered mesh was generated using the Monkhorst-Pack method~\cite{monkhorst1976special}. The k-mesh was re-scaled according to the size and symmetry of each unit cell. For the finite system calculations, a vacuum spacing  of 15A was maintained between periodic unit cells to avoid interaction between the atoms in the adjacent cells~\cite{data}.

\section{Results and Discussion} 

\subsection{Molecular P3HT} 

We begin our analyses by characterizing finite-length P3HT molecules as shown in Fig.~\ref{fig_molecule}(a,b,c). The molecules can have different terminations at each end to maintain a finite stable geometry. In our model, vacuum spacing was introduced in all three directions to ensure the isolation of the molecule and to eliminate spurious interactions due to periodic boundary conditions. To investigate the impact of chain length and terminal groups on the electronic properties of P3HT, two end groups were considered: hydrogen (H) and methyl (CH$_3$), with increasing number of thiophene rings in each configuration. The calculated band gaps of these molecules, extracted from their density of states (DOS), are presented in Table~1.  

\begin{figure}
	\includegraphics[width=8.5cm]{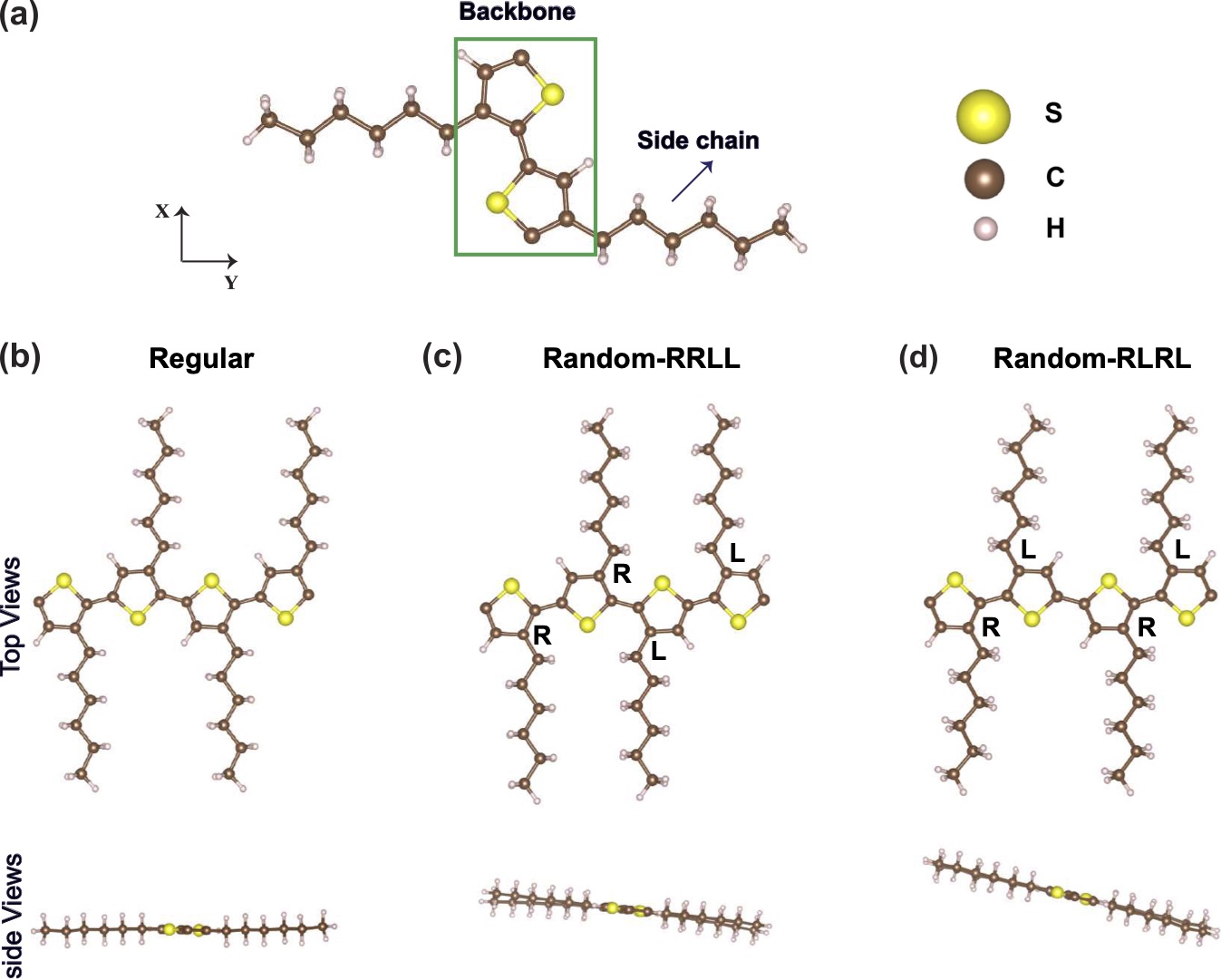}
	\caption{(a) The  isolated unit cell of regular P3HT. The backbone and attached side chains are indicated. (b) Top and side views of the regular and (c-d) two selected random P3HT structures.}
	\label{fig_reg_ran}
\end{figure}

\begin{figure*}
	\centering
	\includegraphics[width=15cm]{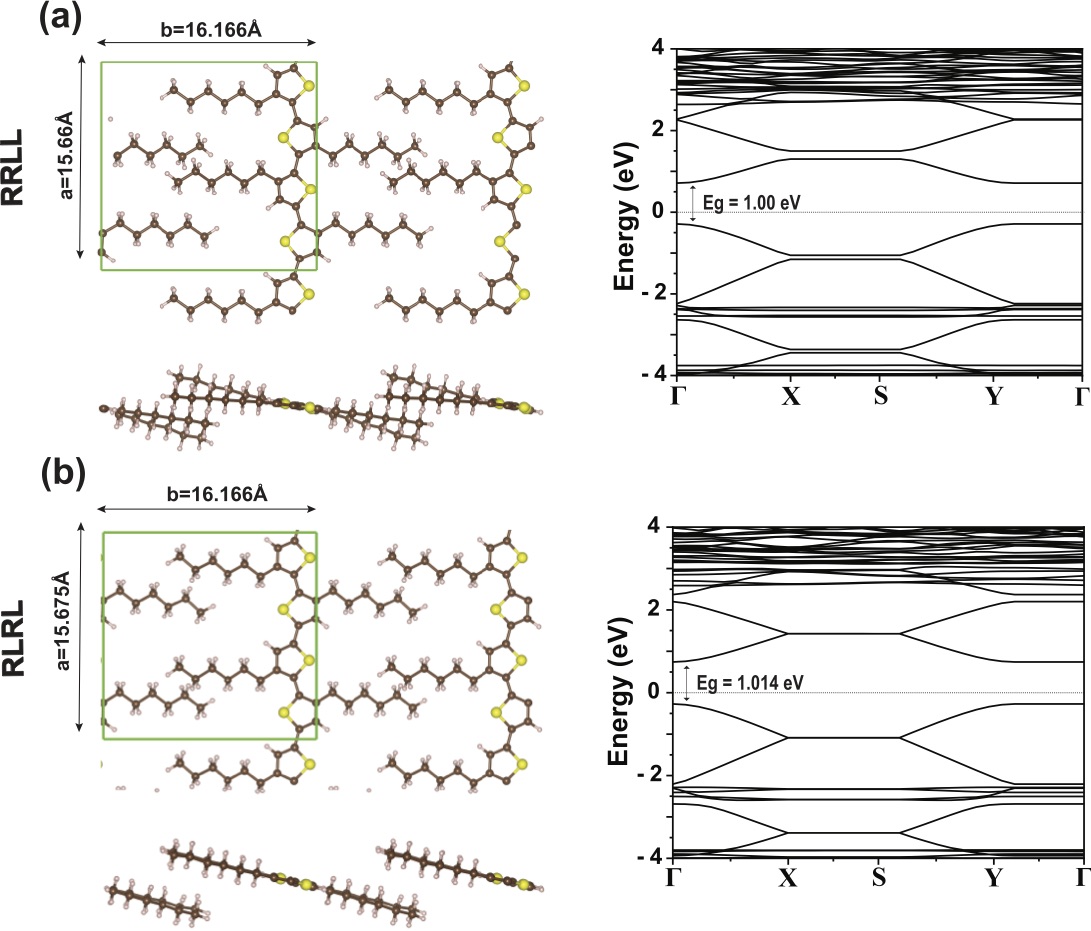}
	\caption{The geometric structures (left) and electronic band diagrams (right) of periodic regio-random P3HT for (a) RRLL and (b) RLRL configurations. The Fermi level is set to zero in all cases.}
	\label{fig_ran}
\end{figure*}

Here, a systematic decrease in the band gap is observed as the molecule size increases, demonstrating the expected behavior of conjugated systems. This behavior is attributed to edge confinement effects in the shorter chains, which limit $\pi$-electron delocalization and increase the energy separation between the Highest Occupied Molecular Orbital (HOMO) and Lowest Unoccupied Molecular Orbital (LUMO) levels.   For H-terminated P3HT molecules, the band gaps are 2.42eV, 1.96eV, and 1.47eV for molecules with 2, 3, and 4 thiophene rings, respectively. A similar trend is observed for CH$_3$-terminated molecules, where the corresponding band gaps are 2.37eV, 1.89eV, and 1.35eV. CH$_3$ termination consistently results in a slight band gap reduction compared to H termination. This behavior indicates that extending the conjugation length enhances electronic delocalization, thereby narrowing the band gap. When P3HT is modeled with periodicity  along the molecule's chain direction, [Fig.~\ref{fig_molecule}(d)], the structure exhibits an even narrower  band gap of approximately 1.08eV, which agrees well with prior results in which a hexyl side chain was substituted by a hydrogen atom \cite{kaloni2015polythiophene}. The increase in band gap when periodicity is restricted in the polymer's chain direction highlights the significant role that side chains play in modulating the electronic structure of P3HT. The trend is consistent with the known electronic behavior of conjugated polymers, where longer chains facilitate better $\pi$-electron delocalization and improved charge transport characteristics \cite{darling2009defect,heeger2001semiconducting,schilinsky2005molecularweight}.

\subsection{Layered P3HT} 

We next merge the molecular P3HT structures to construct a planar film of P3HT. We  examine the standalone P3HT film as a baseline to understand how interaction with graphene influences its electronic and interfacial charge transfer behavior, which we will focus in the next section. The structure of an isolated P3HT monomer is illustrated in Fig.~\ref{fig_reg_ran}(a). As investigated in the previous section, the monomer contains a $\pi$-conjugated thiophene backbone with  two alkyl side chains. By connecting these monomers sequentially, different P3HT polymer chains can be constructed, distinguished by the relative positioning of the alkyl side chains in successive units, as illustrated in Fig.~\ref{fig_reg_ran}(b–d). Based on this molecular arrangement, P3HT can be categorized into two fundamental structural types: (i) regio-regular (ReP3HT), in which all side chains adopt the same orientation along the molecule's backbone, and (ii) regio-random (RaP3HT), where the side chains attach to the thiophene ring at random positions in alternating units. Here, left and right refer to the positions of the thiophene-ring carbon atoms that serve as attachment sites for the alkyl chain.

\begin{figure*}
	\includegraphics[width=15cm]{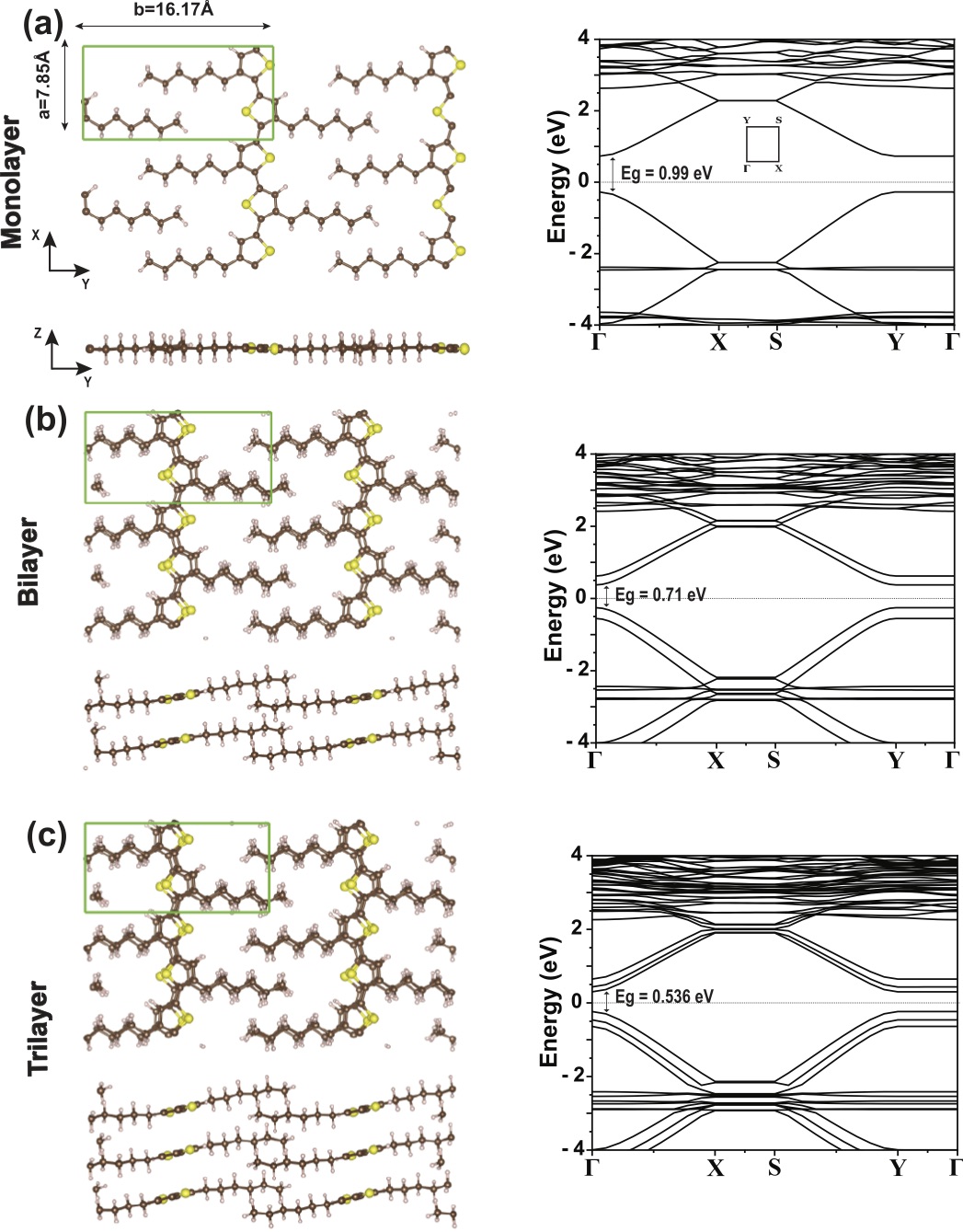}
	\caption{The structures (left) and corresponding electronic band diagrams (right) for (a) monolayer, (b) bilayer, and (c) trilayer periodic regio-regular P3HT. The Fermi level is set to zero. Brown, yellow, and pink spheres represent the C, S, and H atoms, respectively. The rectangular Brillouin zone (BZ) with high-symmetry points is shown.}
	\label{fig_reg_multi}
\end{figure*}

When the P3HT molecules are brought together, they form longer chains and as chains are brought together on the same plane, it is possible to form layered P3HT sheets. Fig.~\ref{fig_reg_multi}(a) shows the relaxed geometry of a repeating unit cell of the P3HT molecule [$(\mathrm{C}_{20}\mathrm{H}_{28}\mathrm{S}_{2})_n$], which consists of two thiophene rings with their alkyl side chains oriented in opposite directions. The optimized chain length along the $y$-direction is found as 7.85~\AA, and the in-plane chain-to-chain distance is 16.17~\AA. These values agree well with previous experimental results of 7.7~\AA~and 16.2~\AA~\cite{prosa1992comb,sirringhaus2000synthetic}. For multilayer P3HT, the calculated interlayer distance is between 3.6  and 4.1~\AA~  which are also consistent with the experimentally measured value of 3.8~\AA~\cite{sirringhaus2000synthetic,brinkmann2006orientation}. Note that in the multilayers, the interlayer distance depends on the point of measurement since the monolayers start to buckle up and down. Also, the stability of the P3HT layers are significantly influenced by their stacking configurations. For free-standing bilayers, we found that AB stacking [Fig.~\ref{fig_reg_multi}(b)] is  energetically favorable in comparison the AA stacking, with an energy difference of approximately 0.38eV per unit cell, consistent with previous studies \cite{kiymaz2016nanowires}. Similarly, for trilayers, ABC stacking [Fig.~\ref{fig_reg_multi}(c)] is more stable than AAA stacking, with an energy difference of  0.68eV. This preferential stacking arises from reduced static hindrance and enhanced $\pi$--$\pi$ interactions, which optimize molecular packing. Consequently, repulsive forces are minimized and van der Waals interactions are enhanced \cite{dag2010packing} in these stacking geometries. Electronic band structure calculations reveal a systematic decrease in the band gap with increasing P3HT layer thickness: from 0.99eV in the monolayer, to 0.71eV in the bilayer, and 0.54eV in the trilayer. This trend is attributed to the strengthening of interlayer interactions mediated by van der Waals forces, leading to increased electronic delocalization and reduced band gaps \cite{cai2014phosphorene}.

\begin{figure*}
	\centering
	\includegraphics[width=15cm]{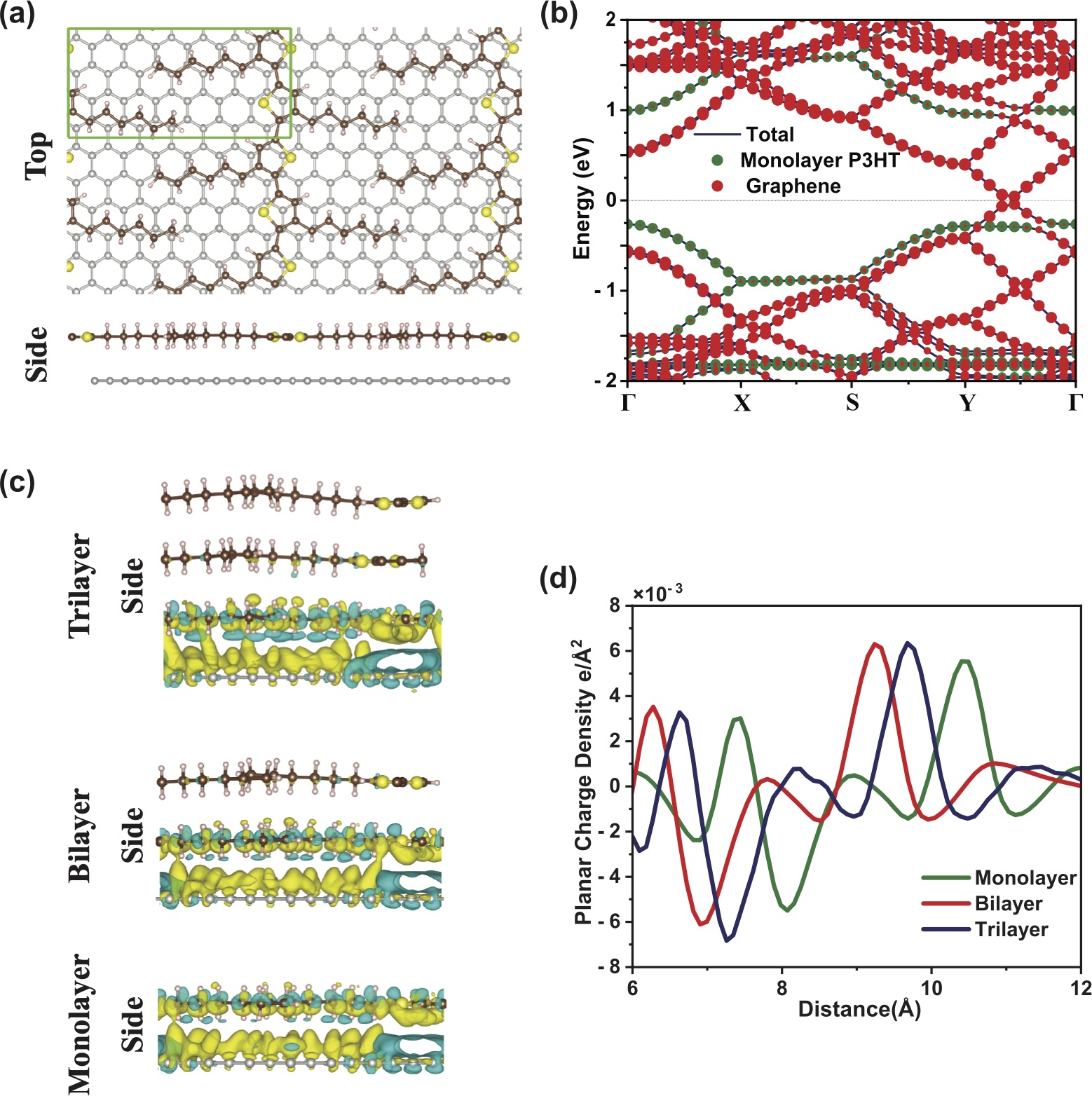}
	\caption{(a) Top (above) and side (below) views of the optimized structure of a monolayer regio-regular P3HT chain deposited on graphene. (b) Projected electronic band structure of the monolayer P3HT/graphene heterostructure, highlighting contributions from  atoms in P3HT (green) and graphene (red) layers. The Fermi level is set to zero. (c) Side views of the charge density difference isosurfaces for monolayer, bilayer, and trilayer P3HT chains on graphene, where yellow and cyan indicate electron accumulation and depletion regions, respectively. (d) Plane-averaged charge density difference profiles along the z-direction corresponding to each configuration.}
	\label{fig_gra_reg}
\end{figure*}

 \begin{figure*}
	\centering
	\includegraphics[width=15cm]{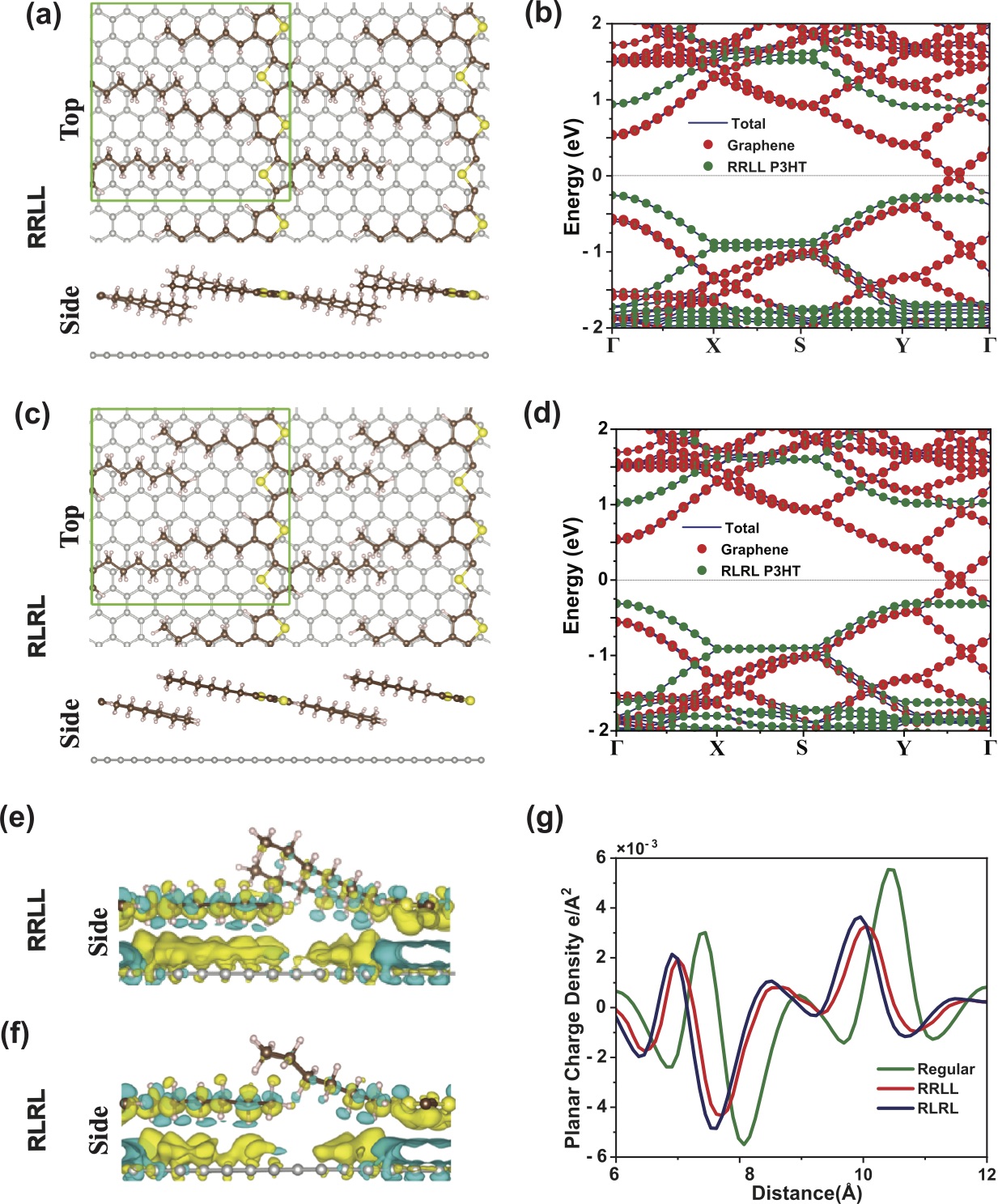}
	\caption{(a, d) Geometrical structures and projected electronic band diagrams of RRLL and RLRL regio-random P3HT chains deposited on a graphene substrate. The Fermi level is set to zero. (e, f) Side views of the charge density difference isosurfaces for the RRLL and RLRL configurations. (g) Comparison of the plane-averaged charge density difference profiles of regular and random P3HT configurations.}
	\label{fig_gra_ran}
\end{figure*}
 
In order to explore the effect of structural disorder on electronic properties, we  chose  two sample random RaP3HT configurations, denoted as RRLL and RLRL, referring to the relative positions of the hexyl side chains on adjacent thiophene rings. In this context, R and L indicate whether the side chain is positioned to the right or left of the polymer backbone, respectively. The relaxed geometries of the repeating unit cells for RRLL and RLRL configurations are shown in Fig.~\ref{fig_ran}(a, b). The optimized chain lengths along the $y$-direction are 15.66A and 15.68A for RRLL and RLRL, respectively, and the in-plane chain-to-chain distances are 16.18A (RRLL) and 16.17A (RLRL). These regiorandom configurations exhibit slightly higher band gaps (up to 1 percent) compared to the ordered structure, due to weaker electronic coupling induced by molecular disorder.

\begin{table}
	\centering
	\caption{
		\label{tab:finite_p3ht_bandgap}
		Band gaps (in eV) of finite P3HT molecules with different end terminations and chain lengths. The band gap decreases with increasing chain length. CH$_3$ termination results in slightly lower band gaps than H termination due to enhanced delocalization effects.}
	\small
	\setlength{\tabcolsep}{6pt}
	\renewcommand{\arraystretch}{1.2}
	\begin{tabular}{cccc} \\
		\hline \hline
		Termination & 2 rings & 3 rings & 4 rings \\
		\hline
		H      & 2.42  & 1.96  & 1.47  \\
		CH$_3$ & 2.365 & 1.89  & 1.353 \\
		\hline \hline
	\end{tabular}
\end{table}

\subsection{Graphene / P3HT Heterostructure}

We next focus on the 2D graphene/P3HT vertical heterostructure to demonstrate the charge transfer across the interface. The optimized structure of a  ReP3HT layer deposited on graphene is illustrated in Fig.~\ref{fig_gra_reg}(a), while the projected band structure highlighting the electronic contributions from both P3HT and graphene is shown in Fig.~\ref{fig_gra_reg}(b). To ensure compatibility with the periodicity of the underlying graphene lattice, the polythiophene backbone was stretched by approximately 3\% compared with its free-standing monolayer counterpart. This adjustment provides a realistic representation of the interface and causes negligible changes in the P3HT band gap. Since the charge transfer mechanism is closely related to the band alignment between the materials \cite{zheng2019heterostructures}, we also analyzed the modulation of the electronic structure of the graphene/P3HT heterostructures.  The charge distribution after the formation of the graphene/P3HT interface was obtained by calculating the difference in the local charge densities, defined as \( \Delta \rho = \rho_{\text{Graphene+P3HT}} - \rho_{\text{Graphene}} - \rho_{\text{P3HT}} \), where each term is computed in the same unit cell. By integrating the charge density difference in the xy-plane the planar-averaged charge density difference was calculated along the direction perpendicular to the interface (z-direction); to further elucidate the spatial distribution of charge transfer across the interface.

Similar calculations were repeated to obtain the electronic properties and charge mechanisms of the graphene/P3HT heterostructures containing bilayer and trilayer P3HT configurations. The binding energy ($E_b$), a key parameter indicating interfacial interaction strength, is calculated as:
\( E_b = E_{\text{Graphene+P3HT}} - E_{\text{Graphene}} - E_{\text{P3HT}} \)
where each term represents the total energy of the combined or isolated system. The computed binding energies increase slightly with the number of P3HT layers: 1.52eV for monolayer, 1.63eV for bilayer, and 1.67eV for trilayer. This suggests that additional P3HT layers mainly enhance intermolecular interactions beyond the first few layers. Band structure analysis shows a decrease in the P3HT band gap from 1.16eV (monolayer) to 0.97eV (bilayer) and 0.86eV (trilayer). Charge redistribution is visualized in Fig.~\ref{fig_gra_reg}(c) and Fig.~\ref{fig_gra_reg}(d), where electrons localize on the graphene side and holes on the P3HT layer. As the number of P3HT layers increases, both charge accumulation and depletion become more pronounced, indicating increased interfacial charge transfer.

\begin{figure*}
	\centering
	\includegraphics[width=15cm]{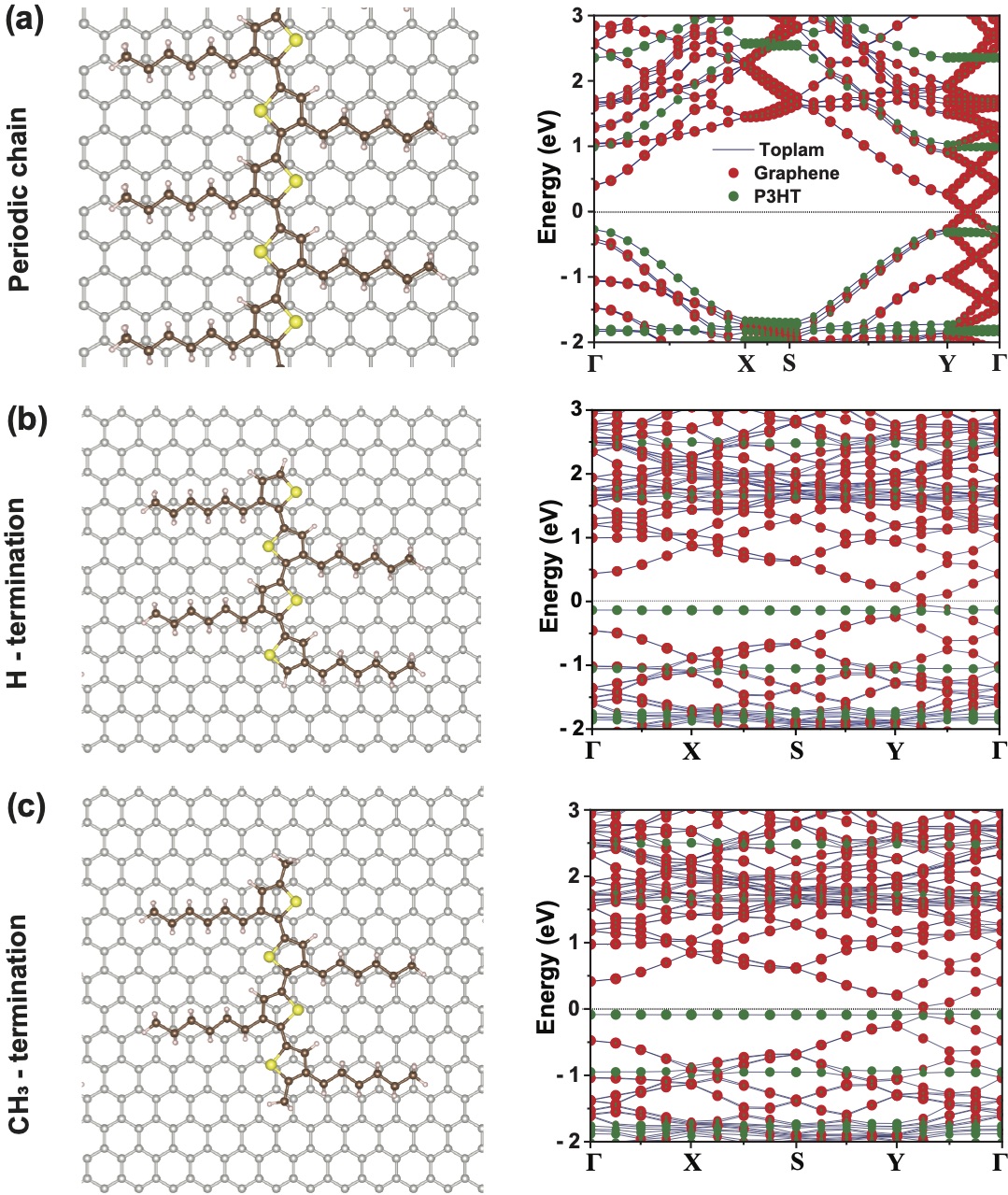}
	\caption{ (Top views of  structure (left) and projected electronic band structure (right) of (a) periodic P3HT chain, (b) finite molecule with CH$_3$ termination, and (c) finite molecule with H termination on graphene. In the band diagrams,  contributions from P3HT (green) and graphene (red) are highlighted. The Fermi level is set to zero.}
	\label{fig_gra_cha}
\end{figure*}

We further analyze the random stacking configurations (RRLL and RLRL) of P3HT on graphene, as shown in Fig.~\ref{fig_gra_ran}(a,c). The binding energy per two thiophene rings is 1.36~eV, which is lower than the regular stacking, correlating with larger band gaps of 1.20eV (RRLL) and 1.34eV (RLRL) as shown in Fig.~\ref{fig_gra_ran}(b,d). Charge density isosurfaces in Fig.~\ref{fig_gra_ran}(e,f) show reduced charge transfer due to the rotated side chains, which reduce interfacial contact. These results agree with earlier observations in metal/P3HT interfaces \cite{ozcelik2020hybrid}, where random configurations hinder $\pi$ - $\pi$ interactions and weaken charge transfer. Further evidence is provided by the planar-averaged charge density difference profiles along the z-direction. As shown in Fig.~\ref{fig_gra_ran}(g), the amplitude of charge density across the graphene P3HT interface is higher in regular stacking than in the random ones, further indicating more efficient charge transport in the former. In contrast, the reduced amplitude in the disordered systems again suggests weaker interfacial coupling and limited electron delocalization.

Here we also note that, when P3HT is modeled as a periodic structure only along the chain direction, the band gap increases to 1.26eV. (Fig.~\ref{fig_gra_cha}(a)). This widening indicates that reduced periodicity weakens electronic coupling, resulting in lower charge transfer efficiency. Side chains thus play a critical role in facilitating interfacial interactions and electronic delocalization.
To further understand edge and length effects, we modeled finite P3HT molecules terminated with either hydrogen (H) or methyl (CH$_3$) groups (Fig.~\ref{fig_gra_cha}(b,c)). Vacuum was introduced in all directions to isolate the molecules. As summarized in Table~2, band gaps decrease systematically as the chain length increases. For H-terminated P3HT, band gaps are 2.78eV (2 rings), 2.18~eV (3 rings), and 1.90eV (4 rings). For CH$_3$-terminated molecules, the band gaps are slightly lower: 2.67eV, 2.10eV, and 1.84eV, respectively. These differences suggest that methyl termination slightly enhances electronic delocalization compared to hydrogen termination.

\begin{table}
    \centering
    \caption{
        \label{tab:bandgaptermination}
        Band Gaps (eV) of Graphene P3HT heterostructures with different terminations and chain lengths.
    }
    \small
    \setlength{\tabcolsep}{6pt}
    \renewcommand{\arraystretch}{1.2}
    \begin{tabular}{cccc} \\
        \hline \hline
        Termination & 2 rings & 3 rings & 4 rings \\
        \hline 
        H      & 2.78 & 2.18 & 1.90 \\
        CH$_3$ & 2.67 & 2.10 & 1.84 \\
        \hline \hline
    \end{tabular}
\end{table}

\begin{figure}
	\centering
	\includegraphics[width=8.5cm]{fig_gra_cha2.jpg}
	\caption{Plane-averaged charge density difference profiles for graphene/P3HT heterostructures with different periodicity and terminations. (a) and (b) show the charge distribution along the out-of-plane direction for H and CH$_3$ terminated finite P3HT molecules, respectively, with different chain lengths (2, 3, and 4 rings). (c) Comparison of the charge transfer characteristics of the periodic two-dimensional layer, periodic one dimensional  chain, and the finite P3HT molecule model with H-termination (4-ring configuration is used for the finite case).}
	\label{fig_gra_cha2}
\end{figure}

Plane averaged charge density difference plots shown in  Fig.~\ref{fig_gra_cha2}(a,b) illustrate enhanced interfacial charge transfer with increasing chain length. Interestingly, the 3-ring system shows the lowest charge density difference. While short chains (2 rings) have stronger edge effects and long chains (4 rings) offer extended $\pi$-conjugation, the 3-ring configuration adopts a structural resonance that stabilizes its frontier orbitals, leading to less pronounced charge transfer. To investigate the influence of periodicity on the interfacial charge transfer in graphene/P3HT heterostructures, the planar charge redistribution for the monolayer, regular P3HT for different configurations is presented in Fig.~\ref{fig_gra_cha2}(c). It can be observed that charge transfer is most effective in the fully periodic configuration, emphasizing the importance of periodicity along both the polymer's chain and hexyle side chains in enhancing interfacial interactions and electronic coupling between P3HT and graphene.

\section{Conclusion}

In this work, we have systematically investigated the electronic properties and charge transfer mechanisms of 2D vertical graphene/P3HT heterostructures using first-principles density functional theory calculations. By exploring a range of structural configurations including monolayer to trilayer P3HT, regular versus random stacking, periodic versus finite molecules, and variations in terminal groups, here we have elucidated the key factors governing the interfacial charge transfer behaviors. Our results reveal that increasing P3HT layer thickness enhances interfacial charge transfer and reduces the band gap, while structural disorder, such as random stacking, significantly weakens electronic coupling. In finite systems, longer molecular chains and hydrogen termination promote more efficient charge redistribution at the interface, underscoring the critical roles of molecular length and end-group chemistry. Charge density difference analysis confirms that P3HT acts as an electron donor, while graphene serves as an electron acceptor across all configurations studied. We provide an in-depth systematic study of ultrathin P3HT layers in contact with graphene, providing valuable insights into the design principles for organic 2D hybrid heterostructures. These insights suggest that controlling molecular ordering, periodicity, and termination can optimize the performance of high-efficiency photovoltaic and optoelectronic devices, such as field-effect transistors and solar cells. For future work, we suggest that defect engineering and the application of external fields and mechanical strain could be useful techniques to modulate and enhance charge transfer in graphene/P3HT hybrid structures.

\begin{acknowledgments}
We acknowledges funding from the Scientific and Technological Research Council of Turkey (TUBITAK) under project no 123F264.  Computational resources have been provided by TUBITAK ULAKBIM, High Performance and Grid Computing Center and by the National Center for High Performance Computing of Turkey (UHeM) under grant numbers 1013432022 and 1019342024.
\end{acknowledgments}

\bibliography{arxiv.bib}

\end{document}